\documentclass{aa}
\usepackage{epsfig}
\usepackage{rotating}
\usepackage{natbib}
\usepackage{graphicx}

\newcommand{\ntt}{{\sl NTT}}
\newcommand{\vlt}{{\sl VLT}}
\newcommand{\vltn}{{\sl Very Large Telescope}}
\newcommand{\hstn}{{\sl Hubble Space Telescope}}
\newcommand{\hst}{{\sl HST}}
\newcommand{\wfpcn}{{\sl Wide Field and Planetary Camera 2}}
\newcommand{\wfpc}{{\sl WFPC2}}
\newcommand{\acsn}{{\sl Advanced  Camera for Surveys}}
\newcommand{\acs}{{\sl ACS}}
\newcommand{\chan}{{\sl Chandra}}

\newcommand{\tmass}{{\sl 2MASS}}

\def \psr{PSR\, B0540$-$69}

\begin{document}
\title{HST/WFPC2 observations of the LMC pulsar PSR\,B0540$-$69 \thanks{Based on observations with the NASA/ESA Hubble Space Telescope, obtained at the Space Telescope Science 
Institute, which is operated by AURA, Inc. under contract No.
NAS 5-26555. }
}

\author{R. P. Mignani\inst{1}
 \and
 A. Sartori\inst{2}
\and
 A. De Luca\inst{3,2}
 \and 
 B. Rudak\inst{4,5}
  \and
 A. S\l{}owikowska\inst{6,7}
 \and 
 G. Kanbach\inst{8}
 \and
 P. A. Caraveo\inst{2}
 }

   \institute{Mullard Space Science Laboratory, University College London, Holmbury St. Mary, Dorking, Surrey, RH5 6NT, UK
                       \and
INAF - Istituto di Astrofisica Spaziale e Fisica Cosmica di Milano,Via Bassini 15, I-20133 Milano, Italy 
      \and
Istituto Universitario di Studi Superiori Viale Lungo Ticino Sforza 56,  Pavia, 27100, Italy
\and
Nicolaus Copernicus Astronomical Center, ul. Rabia\'nska 8, 87100, Toru\'n, Poland
\and
KAA UMK, Gagarina 11, 87-100 Torun, Poland.
\and
Institute of Astronomy, University of Zielona G\'ora, Lubuska 2, 65-265 Zielona G\'ora, Poland
\and
IESL, Foundation for Research and Technology, 71110 Heraklion, Crete, Greece 
\and
Max-Planck Institut f\"ur Extraterrestrische Physik, Giessenbachstrasse 1, 85741 Garching bei MŸnchen, Germany
}
\titlerunning{\hst\ observations of \psr}

\authorrunning{Mignani et al.}
\offprints{R. P. Mignani; rm2@mssl.ucl.ac.uk}

\date{Received ...; accepted ...}

\abstract{
The study of the younger, and brighter, pulsars is important to understand the optical emission properties of isolated neutron stars through observations which, even in the 10m-class telescope era, are much more challenging for older and fainter objects. PSR\, B0540$-$69,  the second brightest ($V\sim 22$) optical pulsar, is obviously a very interesting target for these investigations.}
{The aim of this work is threefold:
constraining the pulsar proper motion  and its velocity on the plane of the sky, 
obtaining a more precise characterisation of the pulsar optical spectral energy distribution (SED) through a consistent set of multi-band, high-resolution, imaging photometry observations
measuring the pulsar optical phase-averaged linear polarisation, for which only a preliminary and uncertain measurement was obtained so far from ground-based observations
.} 
{We performed high-resolution observations of \psr\ with the \wfpcn\ (\wfpc) aboard the \hstn\ (\hst), in both direct imaging and polarimetry modes.}
{From multi-epoch astrometry we set a $3 \sigma$ upper limit of 1 mas yr$^{-1}$ on the pulsar proper motion,  implying a transverse velocity $<250$ km s$^{-1}$ at the 50 kpc  LMC distance. Moreover,  we determined the pulsar absolute position with an  unprecedented accuracy of 70 mas. From multi-band photometry  we characterised the pulsar power-law spectrum and we derived the most accurate measurement of the spectral index ($\alpha_{O} = 0.70 \pm 0.07$) which indicates a spectral turnover between the optical and X-ray bands. Finally, from polarimetry we obtained a new measurement  of the pulsar phase-averaged polarisation degree ($PD =16\%\pm4\%$),  consistent with magnetosphere models depending on the actual intrinsic polarisation degree and depolarisation factor, and we found that the polarisation vector ($22^{\circ} \pm 12^{\circ}$ position angle) is  possibly aligned with the semi-major axis of the pulsar-wind nebula and with the apparent proper motion direction of its bright emission knot.}
{By using the \wfpc\ on the \hst, we performed a comprehensive optical study (astrometry,  photometry, and polarimetry) of  \psr. Deeper studies with the \hst\  can only be possible with the refurbished \acsn\ (\acs) and with the new {\em Wide Field Camera 3} ({\em WFC3}).}


   \maketitle

\section{Introduction}

The pulsar \psr\ in the Large Magellanic Cloud (LMC) is often  referred to  as the  Crab pulsar ``twin''  because of it  is  very  similar in  age  ($\sim 1700$  years), spin period  ($P= 50$  ms),  and  rotational energy loss ($\dot {E}  \sim 10^{38}$ erg~s$^{-1}$). 
\psr\  is the second pulsar discovered in X-rays by the {\em Einstein} Observatory (Seward et al. 1984) and the first extragalactic pulsar detected at any wavelength.  Like the Crab, \psr\ is also embedded in a bright pulsar wind nebula (PWN) visible at wavelengths from the optical to the soft/hard X-rays (e.g. De Luca et al. 2007; Petre et al. 2007; S\l{}owikowska et al. 2007). 
After its discovery, \psr\ has been observed by nearly all X-ray satellites, {\em EXOSAT} (\"Ogelman \& Hasinger 1990), {\em Ginga} (Nagase et al. 1990), {\em ROSAT} (Finley et al. 1993), {\em BeppoSax} (Mineo et al. 1999), {\em Chandra} (Kaaret et al. 2001), {\em Rossi-XTE} (de Plaa et al. 2003), {\em ASCA} (Hirayama et al. 2002), {\em Integral} (S\l{}owikowska et al. 2007), and {\em Swift} (Campana et al. 2008). Its X-ray light curve is characterised by a single, broad peak, very much at variance with, e.g.  that of the Crab pulsar.  In radio, the distance to the  LMC made \psr\  undetectable for a long time  until pulsations  were finally detected from Parkes (Manchester et al. 1993). Giant radio pulses were discovered by Johnston \& Romani (2003),  aligned in  phase with the peak of the X-ray pulse (Johnston et al. 2004), making \psr\ the second youngest radio pulsar to feature this phenomenon. \psr\ is also one of the handful of rotation-powered pulsars with a measured  braking index, obtained from X-ray observations (e.g., Zhang et al. 2001; Cusumano et al. 2003; Livingstone et al. 2005).

In the optical, \psr\ is the second brightest isolated neutron star ($V\sim 22$) identified so far  (see Mignani 2009a,b for recent reviews).   Optical pulsations were detected by Middleditch \& Pennypacker (1985) soon after the X-ray discovery, making \psr\ the third optical pulsar after the Crab (Cocke et al. 1969) and Vela pulsars (Wallace et al. 1977). However, it was only through high-resolution imaging observations with the ESO {\em New Technology Telescope} (\ntt) that its optical counterpart was actually identified (Caraveo et al. 1992; Shearer et al. 1994). The \psr\ optical light curve (Middleditch et al. 1987;  Gouiffes et al. 1992; Boyd et al.  1995; Gradari et al. 2009) is characterized  by a single broad peak, very similar to the X-ray and radio light curves, with a significant dip on top. 
The optical spectral energy distribution (SED) of \psr\ was first measured by Middleditch et al. (1987) from high-speed multi-band photometry which suggested the presence of a possible excess in the U band with respect to an otherwise monotonic power-law continuum ($F_{\nu} \propto \nu^{-\alpha_{O}}$).  A re-analysis of the same multi-band photometry measurements, however, did not yield evidence of the claimed U-band excess (Nasuti et al. 1997). The power-law spectrum demonstrates that the optical emission from \psr\ is of magnetospheric origin, with an optical luminosity consistent with the expectations of the Pacini \& Salvati (1987) model. Low-resolution spectra of \psr\ were obtained by Hill et al. (1997), with the \hstn\ (\hst), and by Serafimovich et al. (2004), with the \vltn\ (\vlt), although the latter was affected by the contamination from the supernova remnant. By using multi-band imaging photometry of \psr\  from archival \hst\ observations, Serafimovich et al. (2004) confirmed that the optical spectrum is dominated by a power-law continuum, although with a spectral index different from the values previously published. They also reported a tentative proper motion measurement for \psr, which, however, was not confirmed by De Luca et al. (2007) using a large time baseline of \hst\ observations. Phase-resolved polarimetry observations of \psr\ were performed by Middleditch et  al.  (1987) but  only yielded an upper limit on the phase-averaged polarisation degree, while image  polarimetry observations of  Chanan \& Helfand (1990) focused on the PWN only.  More recently,  image polarimetry observations of \psr\ were performed  by Wagner \& Seifert  (2000)  using the \vlt.  They reported  a phase-averaged polarisation degree of $\approx$  5\%  (with no associated error) which was probably affected  by  the contribution of the unresolved PWN.  Thus, only  preliminary, or uncertain,  phase-averaged  optical polarisation measurements exist for \psr.

In this paper, we report the results of new astrometry, photometry, and polarimetry measurements of \psr\ performed with the \hst\ as a part of a dedicated programme aimed at the comprehensive study of the pulsar and of its  PWN (Mignani et al., in preparation). From the pulsar astrometry we will constrain its proper motion and its velocity on the plane of the sky, and we will provide an update reference position for future observations. From the pulsar photometry we will measure anew the optical SED, we will study the relation with the X-ray spectrum, and we will verify the presence of a spectral turnover between the two energy bands. From the pulsar polarimetry we will measure  its phase-averaged polarisation and we will test the predictions of different neutron star magnetosphere models. The paper is divided as follows: observations and data analysis are described in Sect. 2, while the results are presented and discussed in Sect. 3 and 4, respectively.

\section{Observations and data analysis}

We observed \psr\  with the {\em Wide Field and Planetary Camera 2} (\wfpc) aboard \hst\ (Programme \#10900, PI Mignani).  
The \wfpc\ is a four-chip CCD detector, sensitive to radiation in  the $1150-11000$ \AA\ spectral range.  In order to exploit the maximum spatial resolution, the pulsar was centred in the {\em Planetary Camera} ({\em PC}) chip which has a pixel scale of $0\farcs045$ and a field-of-view of $35\arcsec \times 35\arcsec$. 
The \wfpc\ observations were performed on June 21st, July 26th, September 25th, and November 5th, 2007 for a total of six spacecraft orbits, corresponding to four different visits. We note that since  our target is in the continuous viewing zone (CVZ),   a declination zone which can be  visible for an entire spacecraft orbit for a few days of the year, our programme could have been carried out in one visit only. However, we decided to split it into different visits since the narrow CVZ time window would have made more difficult to cope with the spacecraft orientation constraints required for the \wfpc\ polarimetry observations (see below). 

\begin{table}
\begin{center}
  \caption{Summary of  the \hst/\wfpc\ observations \psr, with  the  observing dates, the filter names, their pivot wavelength $\lambda$ and bandwidth $\Delta \lambda$ in \AA, and the total integration time T in seconds. }
\begin{tabular}{llcr} \\ \hline
Date  & Filter & $\lambda$ ($\Delta \lambda$) & T  	\\ 
yyyy-mm-dd & &  (\AA) & (s) \\ \hline
2007-06-21    & F336W         &  $3359$  ($204$) & 780 \\
                      & F450W         & $4557$  ($404$) & 780 \\
                      & F555W         & $5442$  ($522$) & 300 \\
                      & F675W         & $6717$  ($368$) & 420 \\
                      & F814W         & $7995$  ($646$) & 600 \\        
                      & F606W/POLQ &  $6038$  ($651$) & 1800 \\
2007-07-26    & F606W/POLQ &            ...                & 1800 \\
2007-09-25	  & F606W/POLQ & 	      ...            &1800 \\
2007-11-05     & F606W/POLQ & 	 ...	    &1800 \\ \hline
2005-11-15     & F555W         & 		...	    & 480 \\
                       & F547M          & $5483$  ($205$) & 1040 \\ \hline
1999-10-17     &  F336W(*)          & ... & 600 \\
                        & F791W(*)          & $7872$ ($519$) & 400 \\
                        & F547M(*)          & ... & 800 \\
                        & F502N           & $5012$ ($27$) & 11000 \\
                        & F673N          & $6732$ ($47$) & 8200 \\ \hline
 1995-10-19     &  F555W(*)        & ... & 600         \\
                         & F656N         & $6564$  ($22$) & 3400 \\
                         & F658N(*)         & $6591$ ($29$) & 4000 \\ \hline
\end{tabular}
\label{data}
\end{center}

(*)  observations used by Serafimovich et al. (2004)

\end{table}

As part of our programme, we performed both multi-band photometry and polarimetry observations (see Table \ref{data}, top section, for a summary).  To maximise the spectral coverage, we observed through the F336W,
F450W, F555W, F675W, and F814W	filters. Observations were carried out in the first visit for two consecutive orbits of the spacecraft.  The last orbit of the first visit was devoted to the first of the four planned polarimetry observations, one at each of the selected polarimetry angle. Indeed, the only possibility to perform polarimetry observations with the \wfpc\ and to keep the target positioned in the same chip, is to rotate the spacecraft to orient the instrument polariser.  This requires that, for each polarimetry angle, the observations must be scheduled in different visits. This means that the choice of the spacecraft roll angle is limited by the guide star/solar panel constraints.  
Polarimetry observations were performed at nominal position angles of $0^{\circ}$, $45^{\circ}$, $90^{\circ}$, and $135^{\circ}$ using the F606W filter 
in combination with the Polariser Quad (POLQ) filter. For each filter, all observations were split into three shorter exposures to filter out cosmic ray hits.  

Additional \wfpc\ observations of \psr, taken through broad-band filters, are available in the \hst\ archive (see Table \ref{data}, lower sections). These include two sets of F555W observations taken on October 19th 1995 (\#6120,  PI: Caraveo) and November 15th 2005  (\#10601, PI: Lundqvist), respectively,   and one set of  F336W and F791W 
observations,  taken on October 17th 1999 (\#7340, PI: Morse).  Observations taken through the medium-band filter F547M 
are available from programmes \#7340 and \#10601.  Finally,  observations taken through the narrow-band filters F656N,  F658N, F502N, 
and F673N 
are available from programmes \#6120 and \#7340. Part of the 1995-1999 \wfpc\ observations were used by Serafimovich et al. (2004) to study the pulsar optical SED. 
To complete the available spectral coverage, we decided to add to our data set the archival observations taken through the F791W filter and, for comparison, those taken through the F336W and F555W  ones. At variance with Serafimovich et al. (2004), we decided not to use the observations taken through narrow-band filters, where the pulsar is fainter and where it is more difficult to isolate its emission from that of the nebula. 

All data were processed on-the-fly at the Space Telescope Science Institute (STScI) through the \wfpc\ {\tt CALWP2} reduction pipeline for bias, dark, flat-field correction, flux, and polarimetry calibration using the closest in time reference calibration frames.  For each filter, we combined single exposures with the {\em STSDAS} task {\em combine} to produce co-added and cosmic-ray free images.

\section{Results}

\subsection{Astrometry}

Firstly, we used the available high-resolution \wfpc\ observations of \psr\ to obtain a compelling constrain on the pulsar proper motion  through optical astrometry.
To do this, we used the long time baseline provided by the observations of  De Luca et al. (2007). In particular, we used all data collected with the F555W and F547M filters (see Table ~\ref{data}) and we considered only the {\em PC} images to maximise the spatial resolution. The selected data set includes five observations spanning $\sim11.7$ years.  In order to measure the \psr\ proper motion through relative astrometry, we applied  the algorithm that we have successfully applied in a number of  previous investigations including, e.g. our first study of the pulsar proper motion (De Luca et al. 2007). We started by selecting a grid of 60 good reference sources  which were detected in the {\em PC} field-of-view and not extended, not saturated, with a high signal-to-noise ratio and not too close  to the CCD edges.   For each source, we computed their pixel coordinates on each image by fitting a 2-D gaussian  to their brightness profile, with an uncertainty of 0.02-0.06 pixels
per coordinate. We evaluated the position of the pulsar optical counterpart in the same  way, with an uncertainty of 0.03-0.04 pixels per coordinate. We corrected the source coordinates for the \wfpc\  geometric distortion (Anderson \& King 2003),  as well as for the ``34$^{th}$ row defect'' (Anderson \& King 1999). We assumed the 1995 image as a reference and we aligned the associated reference grid  along right ascension and declination using the known telescope roll angles. By fitting the reference star positions, we computed the coordinate transformation which yields the best superposition of each frame grid on the 1995 one. In the fitting procedure we discarded ten reference object which yielded large residuals  using an iterative sigma clipping algorithm. The resulting r.m.s. uncertainty on the superposition of the frame grids turned out to be  0.06-0.08 pixel per coordinate (using the remaining 50 reference stars). This coordinate transformation allowed us to convert the  pulsar positions to a common reference frame and to evaluate its displacement in right ascension and declination with respect to its 1995 position. A simple linear fit to the measured displacements as a function of time yields no evidence for any significant proper motion of the pulsar, with a $3\sigma$ upper limit of $0.015$ pixel yr$^{-1}$ per coordinate. Using the well calibrated \wfpc\  plate scale, we set a $3\sigma$ upper limit to the overall pulsar proper motion of 1 mas yr$^{-1}$. Such a result confirms our previous findings (De Luca et al. 2007), but it is slightly more compelling because  of the longer time baseline (11.7 years with respect to $\sim$ 10 years) covered by the used data set. 

Secondly, we used our new \wfpc\ images to compute updated coordinates for \psr\ through optical astrometry.  To this aim, we re-calibrated the image astrometry using as a reference the position of stars selected from the {\em Two Micron All Sky Survey} (\tmass) catalogue (Skrutskie et  al. 2006) which has a mean positional error $\la 0\farcs2$  and  it is linked to the International Celestial Reference Frame (ICRF) with  a $\sim 0\farcs015$ accuracy  (Skrutskie et al.  2006). As discussed in Mignani et al. (2005), \tmass\ is preferred to the {\em Guide Star Catalogue 2.3} ({\em GSC-2.3}; Lasket er at. 2008) and to the {\em U.S. National Observatory B1.0} catalogue ({\em USNO-B1.0}; Monet et al. 2003) for fields in the LMC.   On the other hand,  both the {\em USNO CCD Astrograph Catalogue 2} ({\em UCAC-2}; Zacharias et al. 2004) and  its update {\em UCAC-3} (Zacharias et al. 2009) only provide a sparse astrometric grid in the mosaic \wfpc\ field-of-view.  As a reference for our astrometry re-calibration, we chose the June 2007 image taken through the F555W filter, where  \psr\ is detected with a good signal-to-noise, and for which the correction for the \wfpc\ geometric distortion is well modelled.  We used  the mosaic of the four \wfpc\ chips ($160\arcsec \times 160 \arcsec$ field-of-view; 0\farcs1 pixel scale) since it provides a large enough field-of-view to include a sufficient number of \tmass\ stars to be used for the astrometry re-calibration. We produced the mosaic image with the {\em STSDAS} task {\tt wmosaic} which also applies the correction for the  geometric distortions of the four chips.   
Approximately 60 \tmass\ objects are identified in the mosaic \wfpc\ image.  As done in Sect 3.1, we filtered out extended objects, stars  that are either saturated or  too faint to be used as reliable  astrometric calibrators, or too close  to the CCD edges.   We finally  performed  our astrometric  calibration using  40 suitable \tmass\ reference stars, evenly distributed in the \wfpc\ field-of-view. Then, for both the  \tmass\  stars and the pulsar,  we determined their pixel coordinates in the \wfpc\ reference frame from the centroid of their intensity profile, computed by fitting a 2-D Gaussian. This yielded errors (per coordinate) of $\la 0.01$  and of $\la 0.05$ \wfpc\ pixels on the centroids of the \tmass\ stars and of the pulsar, respectively.   We then computed the fit to the  pixel-to-sky coordinate transformation for the \tmass\ stars using the code {\tt ASTROM}\footnote{http://star-www.rl.ac.uk/Software/software.htm}, based on higher order polynomials. 
By applying the computed fit  to the pixel coordinates of \psr\ we finally obtained: $\alpha_{J2000}=05^h  40^m  11.204^s$, $\delta_{J2000}=  -69^\circ   19\arcmin  54\farcs34$, with an overall accuracy $\delta r = 0\farcs12$ ($1 \sigma$).
We estimated the accuracy on the computed coordinates by adding in quadrature the rms of the astrometric fit, $\sigma r \sim 0\farcs1$,  the uncertainty in the  registration of the \wfpc\ image on the \tmass\ reference frame, $\sigma_{tr}=0\farcs055$\footnote{As in Lattanzi et al (1997),  we defined $\sigma_{tr}=\sqrt 3 \times \sigma_{s} /  \sqrt N_{s}$, where $\sqrt 3$ accounts for  the free parameters in the  astrometric fit  (x-scale, y-scale, and rotation angle) $N_{s}$   is  the   number of \tmass\ stars and $\sigma_{s}=0\farcs2$ is  the conservative mean positional error of  their coordinates (Skrutskie et al.  2006)}, the $\sim 0\farcs015$ accuracy of the link of the  \tmass\  coordinates to the ICRF, and the uncertainty on the pulsar centroid ($\la 0\farcs005$). Since the uncertainty on the reference star centroids  is much smaller than that of the pulsar, we neglected it in our global error budget.  

As a check of our absolute astrometry, we  used each of the  F555W data sets (see Table \ref{data}), which were obtained from observations performed  in different visits, i.e. with different telescope pointings and spacecraft roll angles, to obtain independent measurements of the pulsar position.  Since the pivot wavelength of the F547M filter is essentially the same as that of the F555W one (see Table \ref{data}),  the wavelength dependence of the geometric distortion should not induce any bias when using the correction optimized for the F555W filter (see also De Luca et al. 2007).  Thus, we could also use the two F547M data sets as a reference for our absolute astrometry.  In particular, for the November 2005 observations, we decided to use the F547M data set instead of the F555W one, due to its longer integration time. We note that the other suitable medium/broad-band filter data sets can not be formally used for an independent check of our astrometry since they were obtained from observations performed during the same visits as the F555W and F547M ones and, thus, with the same telescope pointing and roll angles.  In principle, the data sets obtained from the observations performed through the F606W/POLQ filter, which were taken in different visits, could be used, though. However,  possible effects of the polarisation optics on the \wfpc\ astrometry, as well as the mapping of the geometric distortion, are still to be studied, thus making the use of imaging polarimetry data less suitable for astrometry (see also Kaplan et al. 2008).
By re-calibrating the astrometry of each data set through the same procedure described above we obtained: $\alpha_{J2000}=05^h  40^m  11.208^s$,  and $\delta_{J2000}=  -69^\circ   19\arcmin  54\farcs12$ ($\delta r =0\farcs14$; $1 \sigma$) for the October 1995 F555W data set.  Similarly, we obtained:   $\alpha_{J2000}=05^h  40^m  11.193^s$, $\delta_{J2000}=  -69^\circ   19\arcmin  54\farcs12$ ($\delta r =0\farcs15$; $1 \sigma$)    and $\alpha_{J2000}=05^h  40^m  11.206^s$, $\delta_{J2000}=  -69^\circ   19\arcmin  54\farcs10$ ($\delta r =0\farcs13$; $1 \sigma$) for the  November 2005 and for the October 1999 547M data sets, respectively.  We note that all the four sets of coordinates are consistent within $\approx1 \sigma$, which confirms that our first measurement was free of systematics. The marginal, although not significant,  differences in the pulsar coordinates are most likely due to the different number of \tmass\ stars used to compute the astrometric solution and to their relative distribution in the \wfpc\ field-of-view, which depends on the different telescope pointing directions and roll angles. Marginal differences can also be due to time-dependent shifts of the \wfpc\ chips in the \hst\ focal plane, which can affect astrometry of the \wfpc\ mosaics.
We note that the upper limit that we obtained on the \psr\ proper motion (see Sect. 3.1) is much smaller than the error on our absolute astrometry. Thus, we can neglect any displacement of the pulsar in the 11.7 year time span covered by the available \wfpc\ observations and we can compute an even more precise position from the average of the four sets of coordinates independently computed from the F555W/F547M data sets. This yields $\alpha_{J2000}=05^h  40^m  11.202^s$, $\delta_{J2000}=  -69^\circ   19\arcmin  54\farcs17$ ($\delta r =0\farcs07$; $1\sigma$) as the most accurate value of the \psr\ position.

\subsection{Multi-band photometry}

We  used  our  \wfpc\  observations  of \psr\  to  perform  multi-band
photometry  using,   for  the  first  time,  a   set  of  observations
consistently taken at the same epoch (see Table ~\ref{data}).  This is 
important to minimise systematic effects on the pulsar photometry due,
e.g. to  possible long-term variations in the  instrument zero points,
to a degrade in the instrument  efficiency, or to the use of different
calibration data  sets.  In the case of  \psr, high-resolution imaging
photometry observations are, at present, better suited than spectroscopy since they
are  less affected  by  the contamination  from  the bright  supernova
remnant (see also discussion in Serafimovich et al. 2004).

We measured the flux of  \psr\ in our \wfpc\ images through customised
aperture photometry  using the {\em IRAF} package  {\tt digiphot}.  To
maximise  the signal-to-noise  ratio,  we measured  the pulsar  counts
through small  apertures (3  pixel radius) and  we subtracted  the PWN
background measured  in an  annulus of  10 pixel radius  at a  2 pixel
distance from the photometry aperture  not to include the wings of the
point spread  function (PSF). We then applied  the aperture correction
using, per each filter, the specific coefficients given in Holtzman et
al. (1995).  Observations at shorter  wavelengths with the  \wfpc\ are
affected by the presence  of contaminants which progressively build up
on  the CCD  surface  (Bagget  et al.  2002).   The CCD  contamination
results in temporal variations in throughput which cause a decrease in
the measured source flux.  This is an issue for the measurement of the
pulsar flux in  the F336W filter, where the  contamination rate can be
of the order of $\approx 0.06$\% per day and tends to be higher at the
centre  of each  {\em PC}  chip,  where  the pulsar  is positioned.  This
effect is not accounted for by the {\tt CALWP2} pipeline and has to be
corrected   manually.  We   then  applied   to  our   measurement  the
contamination  correction,  computed  during monthly  de-contamination
procedures  of the CCD,  the last  one carried  out about  three weeks
before                                                              our
observations\footnote{http://www.stsci.edu/hst/wfpc2/analysis}.      To
all our flux measurements we then applied correction to compensate for the time and position-dependent charge  transfer efficiency (CTE)  losses of the  \wfpc\ detector,
using the formulae given in Dolphin (2009).

Following the \wfpc\  Data Handbook (Bagget et al.  2002), we computed
the  count-rate  to flux  conversion  using  the  image keywords  {\tt
PHOTFLAM}  and  {\tt PHOTZPT},  respectively,  derived  by the  \wfpc\
photometry   calibration  pipeline.    We   then  derived   magnitudes
$m_{F336W} = 21.77 \pm 0.17$, $m_{F450W} = 22.09 \pm 0.05$, $m_{F555W}
= 22.07 \pm 0.05$, $m_{F675W}  = 22.18 \pm 0.04$, and $m_{F814W}=22.35
\pm   0.03$.   To   complement  our   broad-band   imaging  photometry
measurements, we also  computed the pulsar flux from  the October 1999
F791W  observation (see  Table \ref{data}).  Using the  same approach
described above, we  derived $m_{F791W}= 22.29 \pm 0.04$,  which is in
quite good  agreement with that  obtained through the  slightly redder
F814W filter.   The computed  magnitudes, together with  their errors,
are  summarised in  Table \ref{phot}.  The quoted  formal  errors are
purely statistical and do not account for systematic errors related to
the uncertainty of the \wfpc\ photometry calibration, which are of the
same order of magnitude for all filters.

\begin{table}[h]
\begin{center}
  \caption{Observed \wfpc\ magnitudes of \psr\ and associated errors (in parenthesis).   Magnitudes measured in the present work are listed in the third column while those measured by Serafimovich et al. (2004) are listed in the fourth column for comparison. For the latter, values accounting for the CTE correction are given in brackets. }
\begin{tabular}{llll} \\ \hline
Filter   &  Date  &  \multicolumn{2}{c}{Magnitudes} \\ 
            & yyyy-mm-dd &  \\ \hline
F336W & 2007-06-21 & 21.77 (0.17) &   \\
            & 1999-10-17 & 21.96 (0.19) &22.25 [21.98] (0.13)    \\
F450W & 2007-06-21 & 22.09 (0.05) &  \\
F555W & 2007-06-21 & 22.07 (0.05) &  \\
            & 2005-11-15 & 22.11 (0.04) &  \\
            & 1995-10-19 & 21.99 (0.03) &22.17 [22.14]  \\
F675W & 2007-06-21 & 22.18 (0.04) &   \\
F719W & 1999-10-17 & 22.29 (0.04) & 22.44 [22.37] (0.12)   \\
F814W & 2007-06-21 & 22.35 (0.03) &  \\ \hline
\label{phot}
\end{tabular}
\end{center}
\end{table}

To  have  an independent  cross-check  of  our  measurements, we  also
measured  the  pulsar  flux  from  the  archival  October  1999  F336W
observation  and  from  the  November  2005  and  October  1995  F555W
observations. Not  to introduce systematic effects  in our photometry,
for both filters we used  the same apertures and background areas used
for  the June  2007 observations.   For the  F336W filter  we obtained
$m_{F336W}  = 21.96  \pm 0.19$,  i.e.  very well  consistent with  our
measurement from the June 2007 observation.  For the the F555W filter,
however, we obtained $m_{F555W}= 22.11 \pm 0.04$ and $m_{F555W}= 21.99
\pm 0.03$ for the November 2005 and for the October 1995 observations,
respectively.   Thus, while  the  June 2007  and  November 2005  F555W
measurements  are in  almost perfect  agreement with  each  other, the
October  1995 one shows  a somewhat  larger difference,  between $\sim
-0.08$ and $\sim -0.12$ magnitudes.  This difference, not formally significant though, is larger than the
nominal uncertainty  of 0.02-0.04 magnitues on the  \wfpc\ zero points
(Heyer et al. 2004) and, thus, it implies an actual difference in the measured
CTE-corrected   count-rate,  possibly   related   to  the   on-the-fly
re-calibration of  the \wfpc\  October 1995 F555W  data set.   We also
compared  our   measurements  with  the  pulsar   fluxes  measured  by
Serafimovich et al.  (2004) on the very same  data sets (fourth column
in  Table \ref{phot}), i.e.  for the  October 1995  F555W and  for the
October  1999   F336W  and  F791W  observations.  We   note  that  our
measurements yield  fluxes brighter  by $\approx 0.1-0.2$  magnitudes with
respect  to those  of Serafimovich  et al.  (2004), although still
compatible within the errors. We attribute this difference to the fact
that  they apparently did  not apply  the CTE
correction  to  their measurements, which can be of this order of magnitude.  Indeed, by  applying  such correction  we indeed found  that their measurements  turn out to be  more consistent with ours (see Table \ref{phot}, values in brackets). In particular, their CTE-corrected October 1995 F555W flux measurement  agrees with our June 2007/November 2005 F555W ones, thus confirming that the slightly deviant measurement that we obtained from the former data set might actually reflect a difference in the on-the-fly re-calibration.

\subsection{Polarimetry}

We    measured    the    pulsar    polarisation    using    the    web
interface\footnote{www.stsci.edu/hst/wfpc2/software/wfpc2\_pol\_calib.html}
to  the \wfpc\  polarisation calibration  tool (Biretta  \&  Mc Master
1997).  This program computes the transmission of a polariser element,
for a given filter and  CCD gain, using a synthetic spectrum simulated
by  the {\tt STSDAS}  package {\tt  synphot}. As  a reference  for the
pulsar,   we   assumed    its   power-law   spectrum   determined   in
Sect. 3.2. Then, the program computes the Stokes $I$, $Q$, and $U$ parameters
from the counts measured in  the three images oriented along the three
different   polarisation    angles   and   applies    the   coordinate
transformation to convert the  measured polarisation from the detector
to the sky reference frame using the telescope roll angle information.

We selected  the triplet of  images oriented along the  three position
angles of  0$^{\circ}$, 45$^{\circ}$, and  90$^{\circ}$.  As done  in Sect. 3.2,  for each
image we measured the pulsar counts with the {\em IRAF} {\tt digiphot}
package  using an aperture  of 3  pixel radius  and we  subtracted the
background sampled in  an annulus of 10 pixel  width.  Since the \wfpc\ calibration tool returns the Stokes parameters as a linear combination of the counts measured in the three images (Biretta  \&  Mc Master 1997), this procedures accounts  for  the subtraction  of  the  PWN  contribution (Chanan  \&
Helfand 1990)  to the measured  pulsar polarisation.  We  then applied
the  aperture correction  from  Holtzman  et al.  (1995)  and the  CTE
correction  from Dolphin  (2009) to  the  background-subtracted source
counts. In this way,  we measured the Stokes parameters of the pulsar
$(Q/I)_{PSR}=0.09\% \pm 0.03\%$ and $(U/I)_{PSR}= 0.12\% \pm   0.02\%$. 
These  corresponds  to a phase-averaged polarisation degree
$PD = 15.2 \% \pm 2.5\%$ with the polarisation vector oriented along a
positional angle  $PA =27.7^{\circ}  \pm 10^{\circ}$ (east  from north),
where $PA$ and $PD$ are defined as:

$$PA = [(Q/I)^2 + (U/I)^2]^{1/2} \eqno{(1)} $$

$$ PD  = {1  \over 2} \arctan  (U/Q) \eqno{(2)} $$

The associated  uncertainties on $PD$ and $PA$ are derived from  the statistical errors
on  the  pulsar counts  measured  in each  of  the  three images.  
To  investigate a  possible dependence of  the polarisation degree  on the telescope roll angle  chosen to align the polariser  (see Sect. 2), we repeated our measurement using a different triplet of images taken with a different  sequence  of  roll  angles (45$^{\circ}$, 90$^{\circ}$, and  135$^{\circ}$) but we  found  no  significant difference. 
To the statistical  uncertainty of our measurement we have
to add the uncertainty on  the absolute polarimetry calibration of the
\wfpc,  which  includes  the   effects  of  the  instrument  intrinsic
polarisation. For the {\em PC}  chip, this uncertainty is estimated to
be  of the  order of  $1.5\%$ (Biretta  \& Mc  Master 1997).   We thus
assumed an overall error of  $3\%$ on the measured polarisation degree
of the pulsar.

Of course,  this value must be  corrected for the  contribution of the
foreground polarisation towards \psr, which is produced by the dust in the
interstellar medium (ISM) and by the integrated components of both the
Galaxy  and  the LMC  and  it is  not  negligible,  in principle.   To
estimate the contribution of  the foreground polarisation, we measured
the  polarisation   degree  of  a  number  of   test  stars  uniformly
distributed across  the {\em PC} field -of-view. We  followed the same
recipe  used for  the pulsar  but through  an automatic  procedure for
source detection  and count measurement  in each of the  three images,
aperture and CTE   corrections,   source  list   matching,   and  Stokes   parameter
computation through the \wfpc\  polarisation calibration  tool. For the test stars we assumed a flat spectrum to compute the transmission of the POLQ polariser,  which is accurate within $\sim  1\%$ as first approximation (Biretta  \& Mc  Master 1997).   From the
starting  sample  of test  stars  we  then  selected those  which  are
brighter than $m_{F555W} = 22$ and  which can thus be used as a better
reference for our  estimate.  From these stars we  then selected those
($\sim 50$) for which the  polarisation degree $PD$ is measured with a
significance  of at  least $3  \sigma$.  After  applying a  $1 \sigma$
clipping filter, we  ended up with a working sample  of $\sim 20$ test
stars   from   which  we   derived   the   average  Stoke   parameters
$(Q/I)_{stars}=-0.029\%$  (0.023\% rms)  and  $(U/I)_{stars}= 0.015\%$
(0.029\% rms).  From these  values we computed an average polarisation
degree $PD_{stars} = 3.3\% \pm  2.5\%$ and an average a position angle
$PA_{stars} \approx 75^{\circ}$.  After accounting  for the
$\approx 1.5\%$ uncertainty on the absolute polarimetry calibration of
the \wfpc, we then ended up  with $PD_{stars} = 3.5\% \pm 3\%$, which we
assumed as an  estimate of the foreground polarisation.   We note that
our value is compatible with  the $\approx 2-3\%$ estimated by Chanan \&
Helfand (1990) using a larger sample of test stars.  After subtracting
the  foreground  polarisation  in  the  Stokes  parameter  space,  the
corrected Stokes parameters of the pulsar are then $(Q/I)_{PSR}=0.12\%
\pm  0.03\%$ and  $(U/I)_{PSR}=  0.11\% \pm  0.03\%$.  From these,  we
finally computed the intrinsic polarisation degree of the pulsar $PD =
16 \% \pm 4\%$ and the polarisation position angle $PA=22 ^{\circ} \pm 12^{\circ}$,
where  the attached  errors also  account for  the rms  on  the Stokes
parameters of the stars used to estimate the foreground polarisation.

\section{Discussion}

\subsection{The pulsar astrometry}

Unfortunately, the lack  of a measurable proper motion  for \psr\ does
not  allow  to search  for  possible  connections  between the  pulsar
kinematics, its polarisation properties, and the PWN morphology, which
have been found, e.g. for the Crab and Vela pulsars (see Mignani 2009c
and references therein).  The computed $3 \sigma$ upper limit of 1 mas
yr$^{-1}$  (see  Sect.  3.1)  on   the  \psr\  proper  motion  sets  a
corresponding upper limit of $\sim  250$ km s$^{-1}$ on its transverse
velocity.  Although  this value is lower  than the peak  of the pulsar
transverse velocity  distribution (e.g., Hobbs  et al. 2005),  it does
not  allow to claim  a peculiarly  low velocity  for \psr\  unless one
lower   the  constraints   on   the  measured   upper   limit  to   $1
\sigma$. Indeed,  several pulsars feature  transverse velocities lower
than $100$  km s$^{-1}$  including, e.g. the  Vela pulsar for  which a
velocity  of  $\sim  65$  km~s$^{-1}$  has been  inferred  from  \hst\
astrometry (Caraveo et al. 2001).

\begin{table}
\begin{center}
  \caption{Compilation of the \psr\ coordinates available from the literature and associated errors  $\Delta \alpha$ and $\Delta \delta$.  }
\begin{tabular}{llllr} \\ \hline
$\alpha_{J2000} ^{(hms)}$ & $\Delta \alpha ^{(s)}$ &  $\delta_{J2000}^{(\circ ~'~")}$ & $\Delta \delta ^{(")}$ &  Source \\ \\\hline
05 40 11.03   &  0.40  & -69 19 57.5    & 2.0 & X-rays (1) \\	                           
05 40 11.16   &  0.15  & -69 19 57.79  & 0.90 & Optical (2) \\  			
05 40 10.980 & 0.090 & -69 19 55.17  & 0.50 & Optical (3) \\                           
05 40 11.221 & 0.090 & -69 19 54.98  & 0.50 & X-rays (4) \\                            
05 40 11.173 & 0.120 & -69 19 54.41  & 0.70 & Optical (5) \\                         
05 40 11.160 & 0.040 & -69 19 53.90  & 0.20  & X-rays (6)    \\  \hline             
05 40 11.202 & 0.009 & -69 19 54.17 & 0.05 & Optical (7) \\ \hline
\label{coo}
\end{tabular}
\end{center}
(1) Seward et al. (1984); 
(2) Manchester \& Peterson (1989);
(3) Caraveo et al. (1992); 
(4) Kaaret et al. (2001); 
(5) Serafimovich et al. (2004); 
(6) Livingstone et al. (2005); 
(7) this work
\end{table}

The  computed  average  coordinates   of  \psr,  together  with  their
associated errors, are listed  in Table \ref{coo}, for comparison with
those  obtained from  previous  works.  After  its  discovery by  {\em
Einstein} (Seward et al. 1984),  the coordinates of \psr\ were firstly
revised  by  Manchester  \&   Peterson  (1989),  from  optical  timing
observations, and later by Caraveo  et al. (1992), from the astrometry
of    the   optical    counterpart   identified    in    their   \ntt\
images.  We note that   the  discovery   of  the  pulsar   in  radio
(Manchester et al. 1993) did  not allow to obtain  more precise
coordinates. Indeed,  its very low radio flux density at the LMC  distance, $S_{400}= 0.7$ $\mu$Jy,  makes it difficult to obtain an accurate radio timing position. Thus,  the optical coordinates  of Caraveo et  al. (1992)
were thereafter assumed as a  reference. In particular, they were used
to compute the  pulsar X-ray and radio timing  solution by, e.g. Zhang
et al. (2001), Cusumano et al. (2003) but, surprisingly enough, not by
Finley  et   al.  (1993)  and   Manchester  et  al.   (1993).   \chan\
observations (Kaaret  et al. 2001) yielded an  updated X-ray position. The former,
and  actually not  the latter  as  the authors  claim, was  used as  a
reference by Johnston et al.  (2004) for the pulsar X-ray/radio timing
solution, and  by de Plaa et  al. (2003).  A new  optical position was
obtained by Serafimovich  et al. (2004) from \hst\  data, while  a new  X-ray
position  was derived  by  Livingstone  et al.  (2005)  based on  {\em
Rossi-XTE} timing observations.

\begin{figure}
\centering
\includegraphics[height=8.5cm,angle=0,clip]{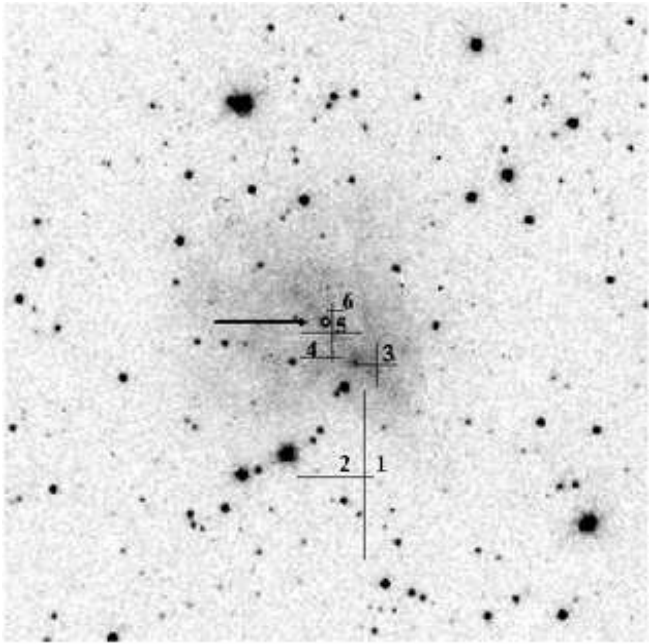}
  \caption{$15\arcsec \times 15\arcsec$ region of the \psr\ field observed with the \hst/\wfpc\ in the F555W filter (October 1995 observation). North to the top, East to the left. The crosses mark the different positions of the pulsar obtained from previous X-ray and optical observations (black) and from our \wfpc\ observation (white). The cross arms correspond to the $1 \sigma$ error in each coordinate. Different positions are numbered according to the reference publications (see Table \ref{coo}). The pulsar is indicated by the arrow for clarity. }
\label{astro}       
\end{figure}

To visualise  the difference and the improvement  on the determination
of the \psr\ position, we plotted all coordinates from Table \ref{coo}
on   the   October   1995    \wfpc\   F555W   image   of   the   field
(Fig. \ref{astro}).  We see  that the original pulsar coordinates from
Seward  et al.  (1984) and  from  Manchester \&  Peterson (1989)  fall
$\approx 3\farcs5$ away from the pulsar position and, actually, out of
the nebula  shell. On  the other hand,  the coordinates of  Caraveo et
al. (1992) and  Kaaret et al. (2001) are indeed  within the nebula but
they are clearly  offset from the actual pulsar  position, while those
of Serafimovich et  al. (2004) and Livingstone et  al. (2005) are more
consistent  with the pulsar  position, although  the former  have much
larger errors.   Thus, thanks to  the sharp angular resolution  of the
\wfpc\ and to the high astrometric accuracy of \tmass, our coordinates
are the most  accurate obtained so far, providing a  factor of $\ga 4$
improvement with  respect to the  most recently published  values.  In
particular,  our   \wfpc\  coordinates  supersede   both  the  optical
coordinates of Caraveo et al. (1992), which were obtained using images
with lower spatial resolution  (0\farcs13/pixel) and the {\em GSC 1.0}
($\sigma_{s}\sim1\farcs0$;  Lasker   et  al.  1990)   as  a  reference
catalogue,  and  those  of  Serafimovich  et al.  (2004),  which  were
obtained using an  early release of the {\em GSC 2.0},  as well as the
X-ray  coordinates  of  Kaaret   et  al.  (2001)  and  Livingstone  et
al. (2005), obtained through \chan\ imaging and {\em Rossi-XTE} timing
observations, respectively.

The better  determination of the \psr\ absolute  position is important
for follow-up non-imaging observations.   In the case of, e.g. spectroscopy,
it will make  it possible to accurately centre the  pulsar in the slit
when  blind   presets  are  made  necessary,   e.g.  for  ground-based
observations due  to the difficulties in resolving  the pulsar through
lower spatial resolution  and seeing-dominated acquisition images.  As
a consequence, this  will allow to use narrower  slits without causing
any loss of signal from the  pulsar. This is important to minimise the
background  contamination from  the supernova  remnant  which hampered
ground-based  spectroscopic   observations  performed  so   far  (see,
e.g. Serafimovich  et al.  2004).

\subsection{The pulsar spectrum}

\begin{figure}
\centering
 \includegraphics[height=8.5cm,bb=20 180 560 720, angle=0,clip]{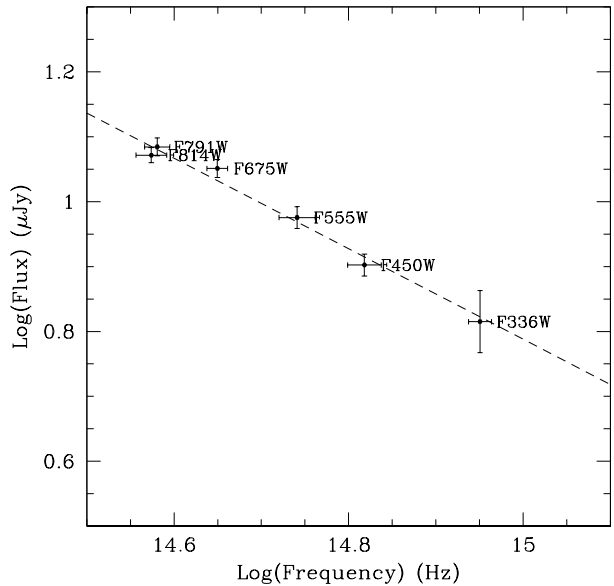}
  \caption{Optical spectral energy distribution of \psr\ derived from the available multi-band \wfpc\ photometry (Table \ref{data}). Points are labelled according to the filter names. The dashed line is to the best fit power-law spectrum.  }
\label{spec}       
\end{figure}

\begin{figure}
\centering
 \includegraphics[height=9cm,angle=270,clip]{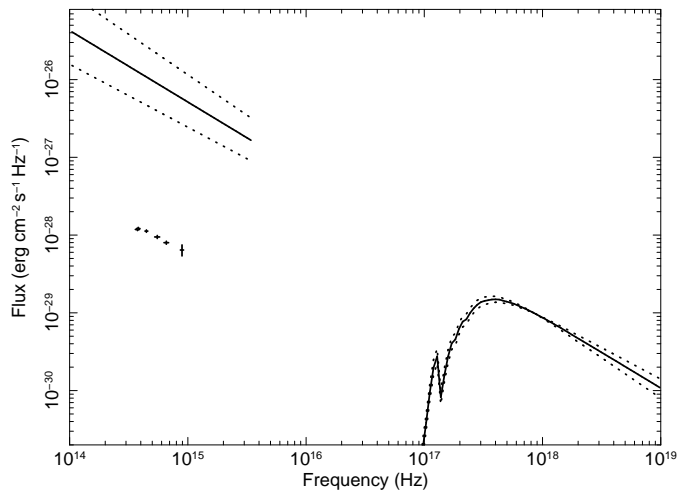}
  \caption{Optical spectral energy distribution of \psr\  (points) compared with the power-law model (Kaaret et al. 2001) best-fitting the \chan\ X-ray spectrum (solid line) and its extrapolation in the optical domain. Dashed lines correspond to a $1 \sigma$ uncertainty on the model parameters. }
\label{multispec}       
\end{figure}

We  used   our  multi-band  \wfpc\  photometry  of   \psr\  to  better
investigate its  optical SED.  For  consistency  with previous  works,  we  computed the  interstellar extinction towards the pulsar using  as a reference an $E(B-V)=0.2$ and
$R=3.1$,  which have been verified  by  several independent measurements (see Serafimovich et  al. 2004 and references therein). We derived the  extinction coefficients in the \wfpc\ filters
from the extinction  curves of Fitzpatrick (1999).  The  pulsar SED is
shown  in  Fig.  \ref{spec},  after correction  for  the  interstellar
extinction. The  plotted errors on the spectral fluxes  are purely statistical
and do not account for the systematic 0.02-0.04 magnitude error on the
\wfpc\  zero points  (Heyer et  al. 2004).  A linear  fit to  the data
points yields a spectral index $\alpha_{O}=0.70\pm 0.07$.  As seen, our
fit clearly confirms that there is no evidence for the "U-band excess" originally claimed
by Middleditch et al. (1985).

We  compared our  best-fit power-law  spectrum with  that  measured by Serafimovich et al. (2004).  They derived a spectral index $\alpha_{O}
= 1.07^{+0.20}_{-0.19}$  which is somewhat steeper  than that measured
by us,  although not formally  incompatible.  As we stated  above, the
power-law slope is  only marginally affected by the  difference in the
assumed  interstellar  extinction  correction.  Indeed,  applying  the
extinction  coefficients of  Cardelli et  al. (1989)  to  our spectral
fluxes,  as done by  Serafimovich et  al. (2004),  yields a  power law
slope  which is  substantially identical  to that  obtained  using the
extinction coefficients  of Fitzpatrick (1999).   Thus, the difference
in  the power-law  slope is  most likely  intrinsic to  the photometry
measurements. First of all, Serafimovich  et al. (2004) fitted a lower
number of points (see Table  \ref{data}) and used the flux measured in
the  medium/narrow-band F547M  and F658N  filters, while  we  fitted a
larger  number  of  points and  we  only  used  the flux  measured  in
wide-band filters.  Moreover,  while we used  similar photometry
apertures,  we applied  aperture corrections  specific for  the \wfpc,
which are  wavelength-dependent (Holtzman et al. 1995),  and we applied
the CTE correction which, apparently,  was  not accounted for by Serafimovich  et al. (2004).

The updated  value of the  \psr\ optical spectral index  is now
more consistent  with those of the other  optically identified pulsars
which  all feature  relatively steep  spectral indexes  ($\alpha_{O} \ga
0.4$),  while only  the Crab  and Vela  pulsar feature  a  nearly flat
power-law spectrum  (see Mignani  et al. 2007a  for a  comparison).  In particular,  \psr\ may be the  pulsar  with  the steepest  optical spectral index.  
We also  note that the
new value  of $\alpha_{O}=0.70 \pm 0.07$  is very close to  that of the
X-ray spectral  index $\alpha_{X}=0.83 \pm 0.13$  measured with \chan\
(e.g.,  Kaaret  et  al.  2001). Interestingly,  however,  the  optical
spectrum is not the continuation  of the X-ray one.  Indeed, the \hst\
spectral  fluxes  lie about  a  factor  of  100 below  the  low-energy
extrapolation of the \chan\ 0.1-10 keV X-ray spectrum (see Fig. \ref{multispec}).
This difference can not
be accounted for by  possible problems in the subtraction of
the PWN background or by other kinds of systematic effects, and 
confirms the presence  of a spectral turnover between  the optical and
the X-ray bands (see also Fig. 14 of Serafimovich et al. 2004).

\begin{figure*}[ht]
\centering
\includegraphics[height=6.5cm,angle=270,clip]{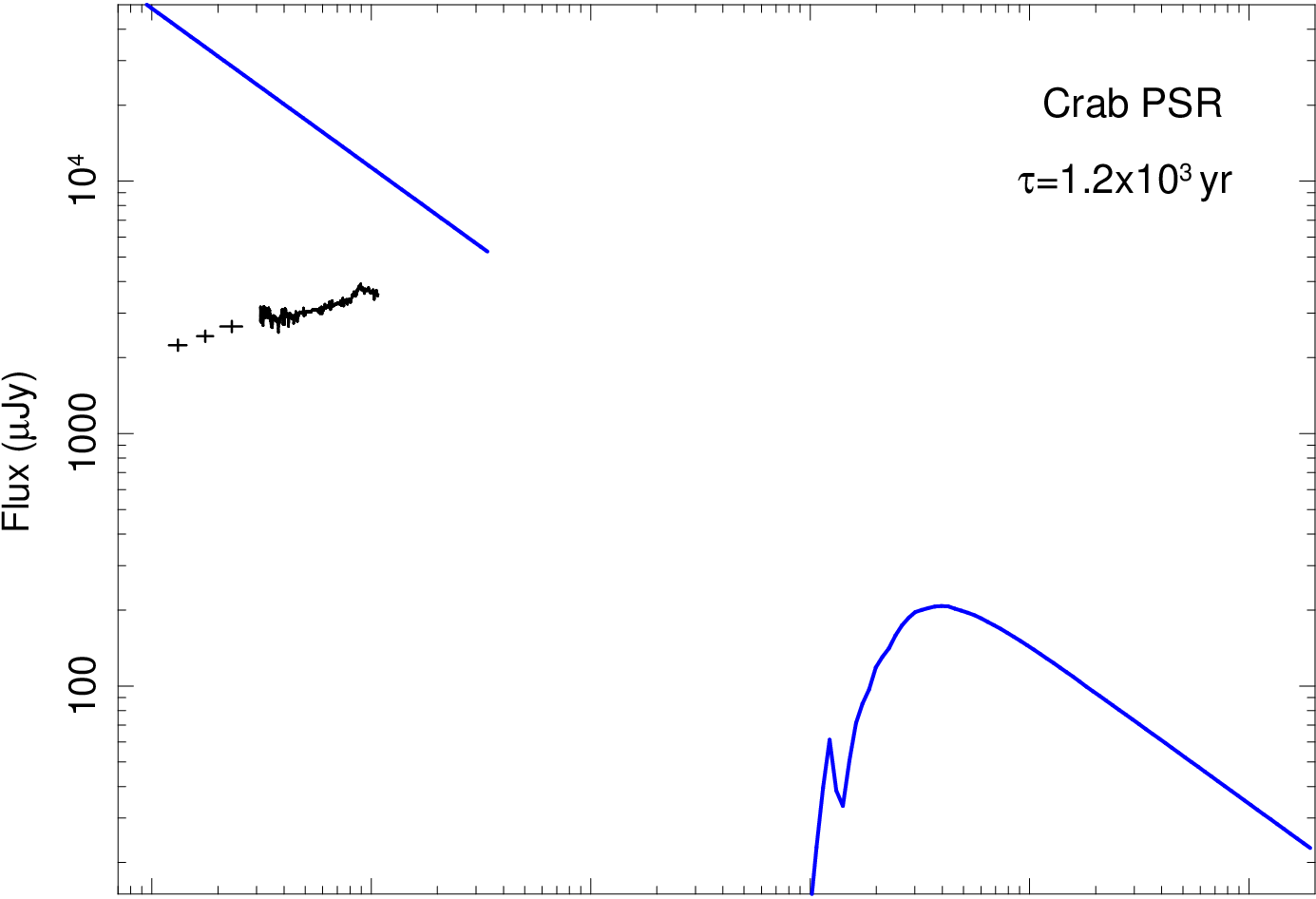}
\includegraphics[height=6.5cm,angle=270,clip]{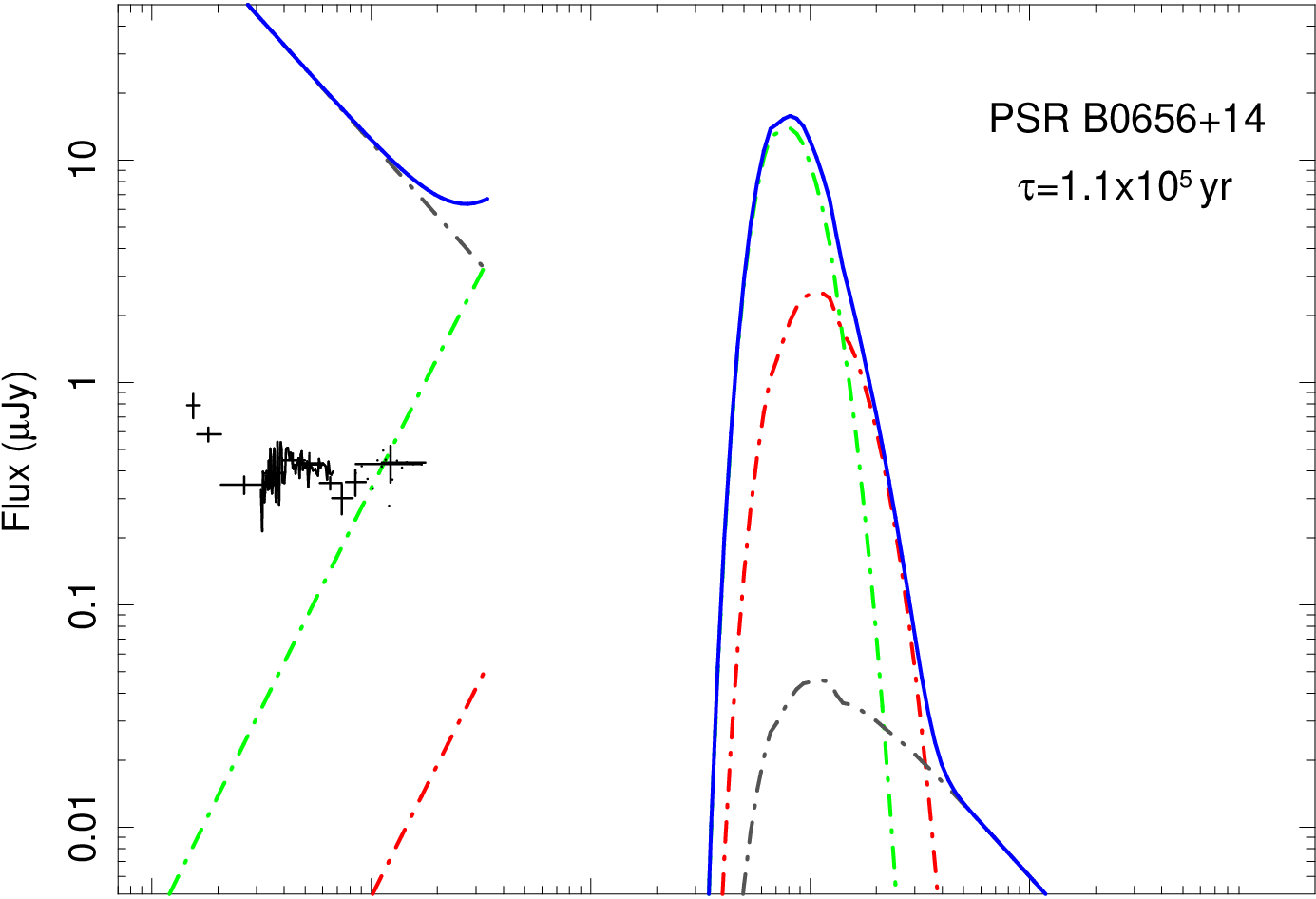}
\includegraphics[height=6.5cm,angle=270,clip]{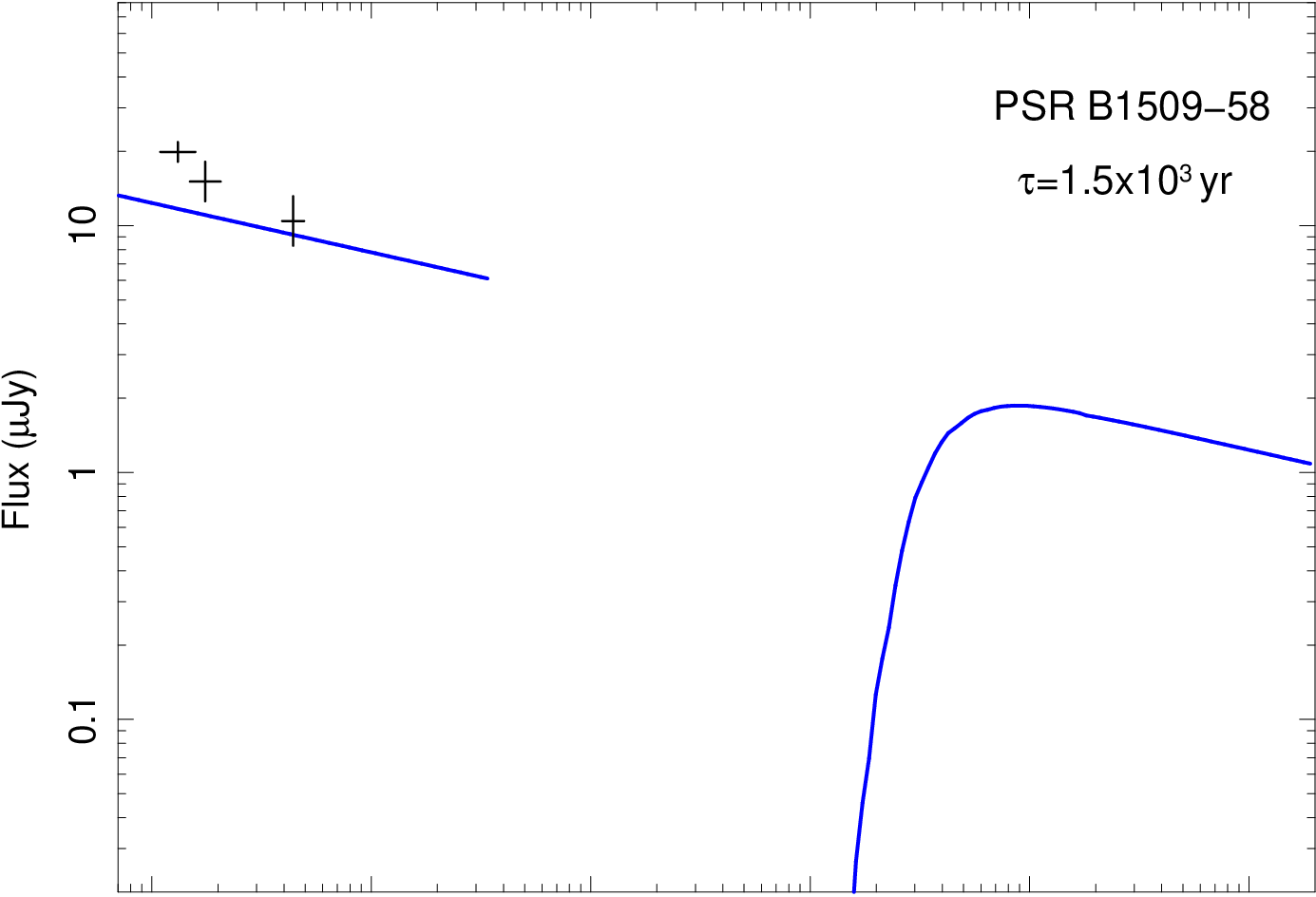}
\includegraphics[height=6.5cm,angle=270,clip]{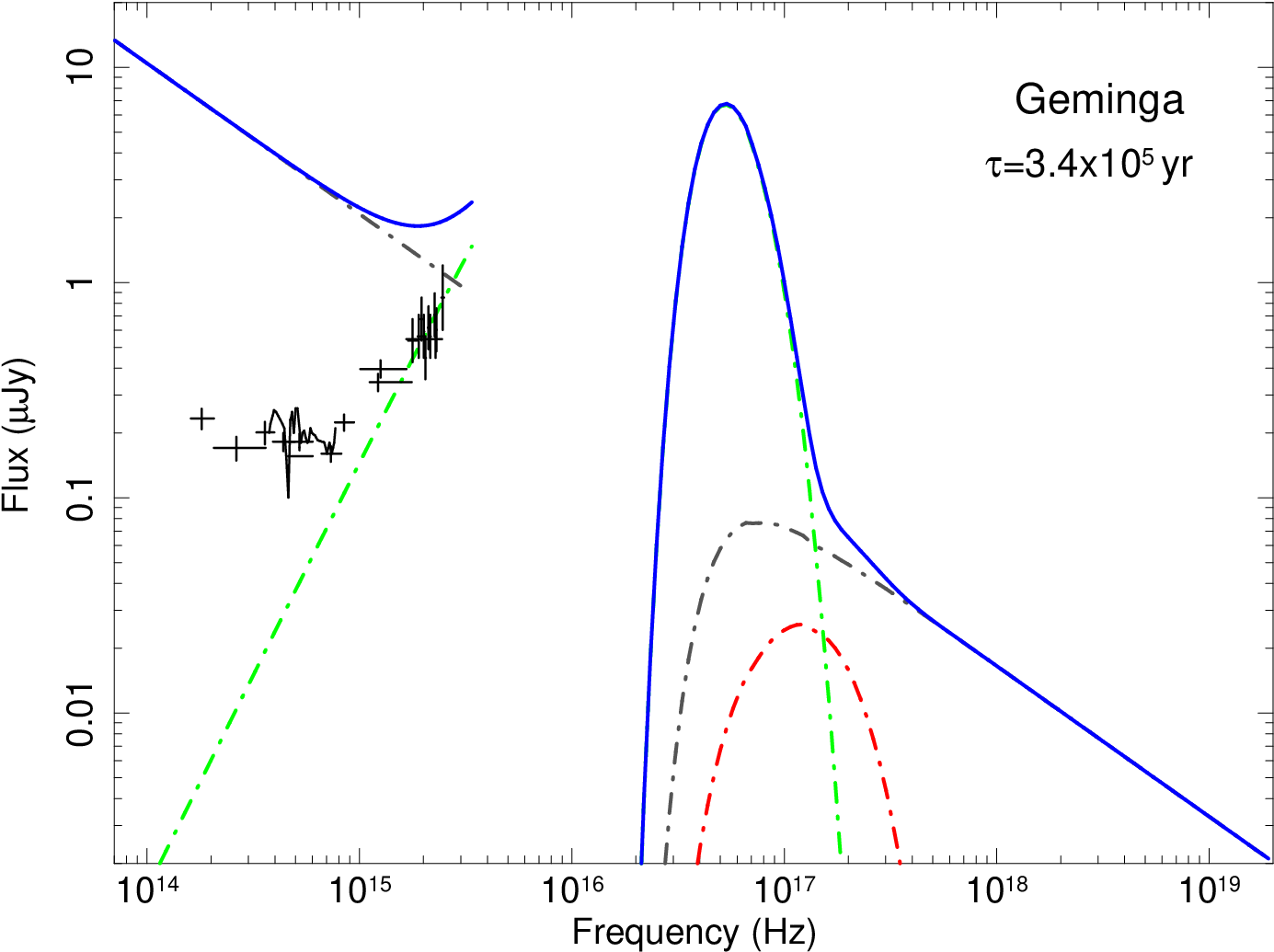}
\includegraphics[height=6.5cm,angle=270,clip]{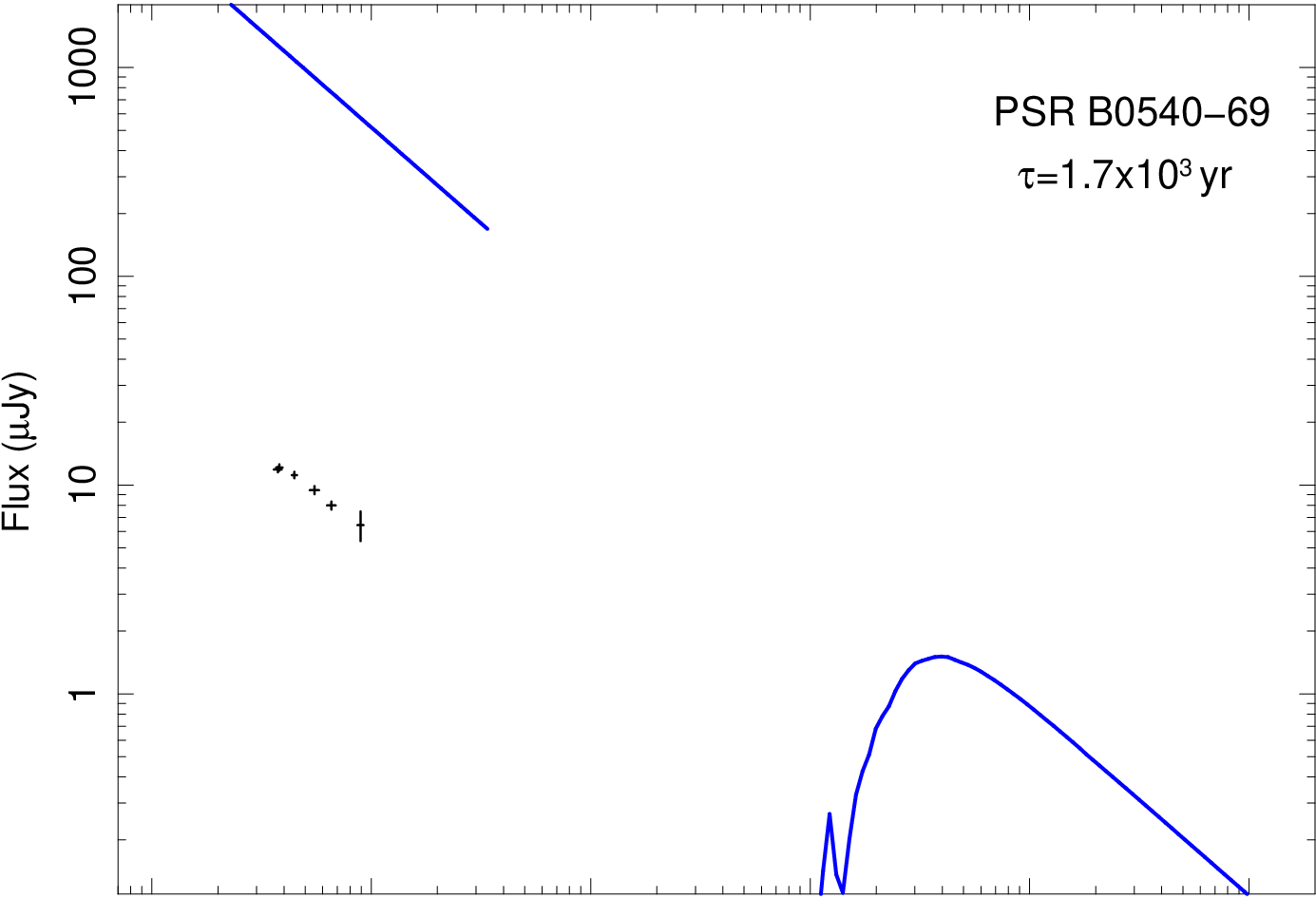}
\includegraphics[height=6.5cm,angle=270,clip]{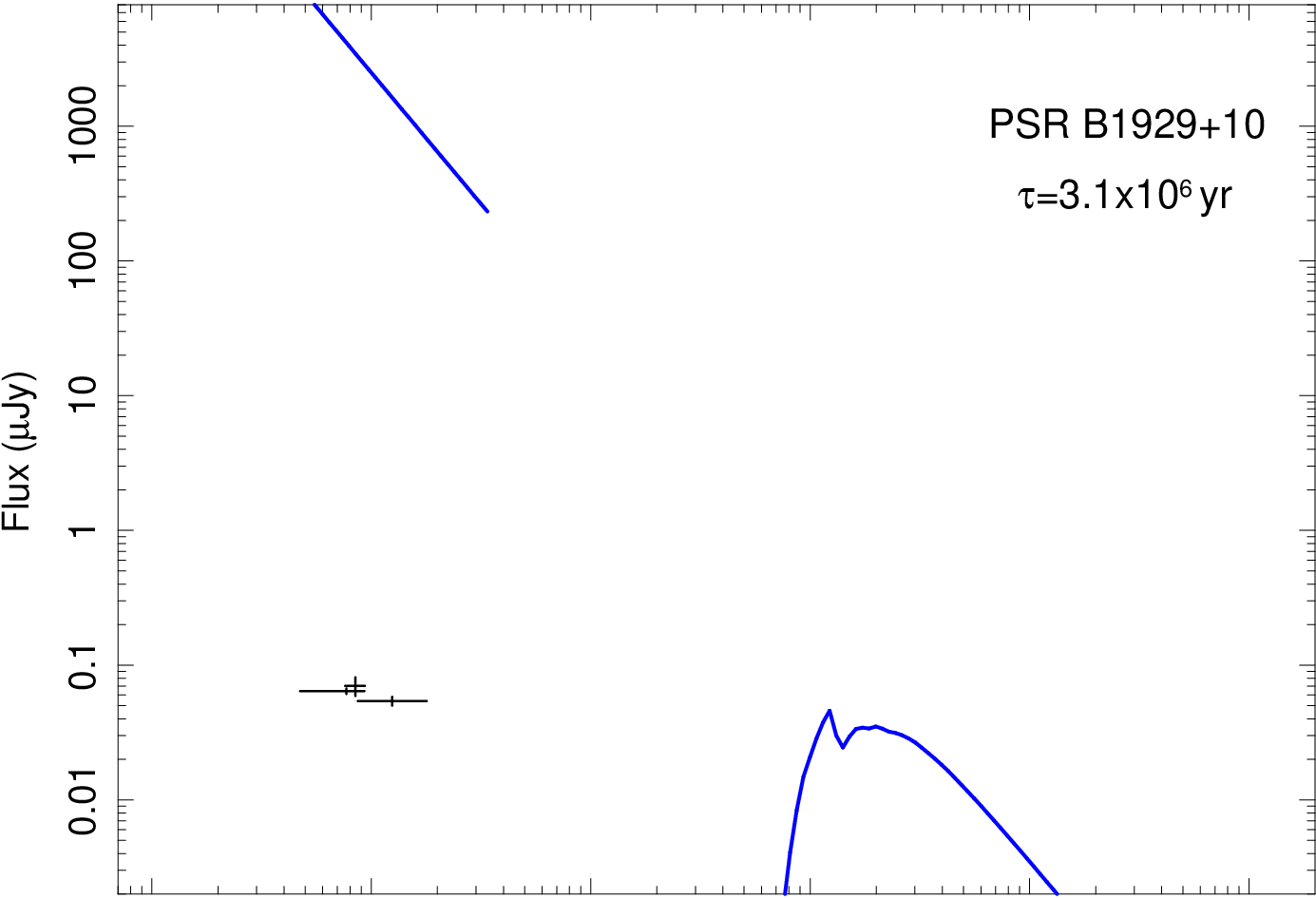}
\includegraphics[height=6.5cm,angle=270,clip]{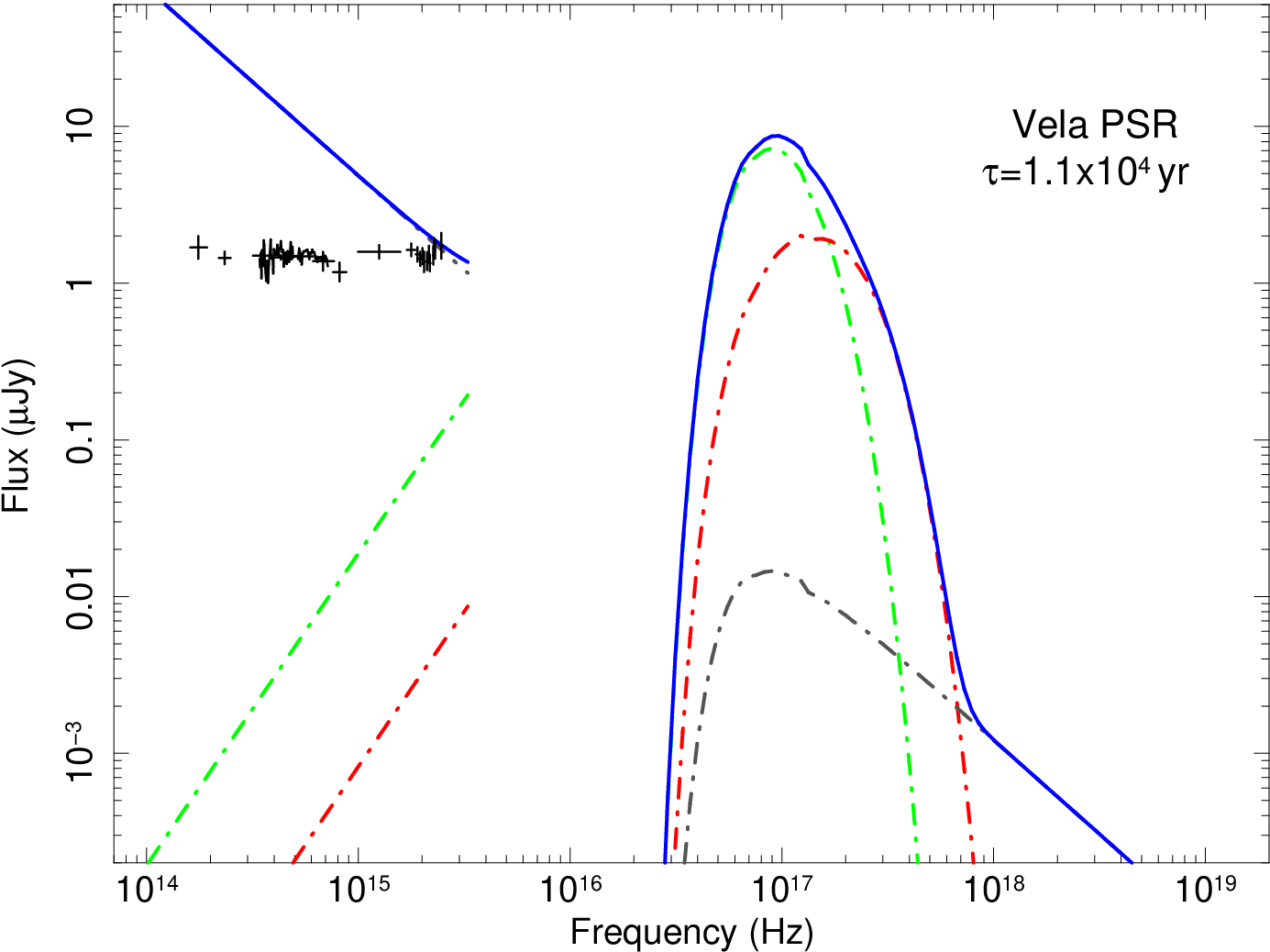}
\includegraphics[height=6.5cm,angle=270,clip]{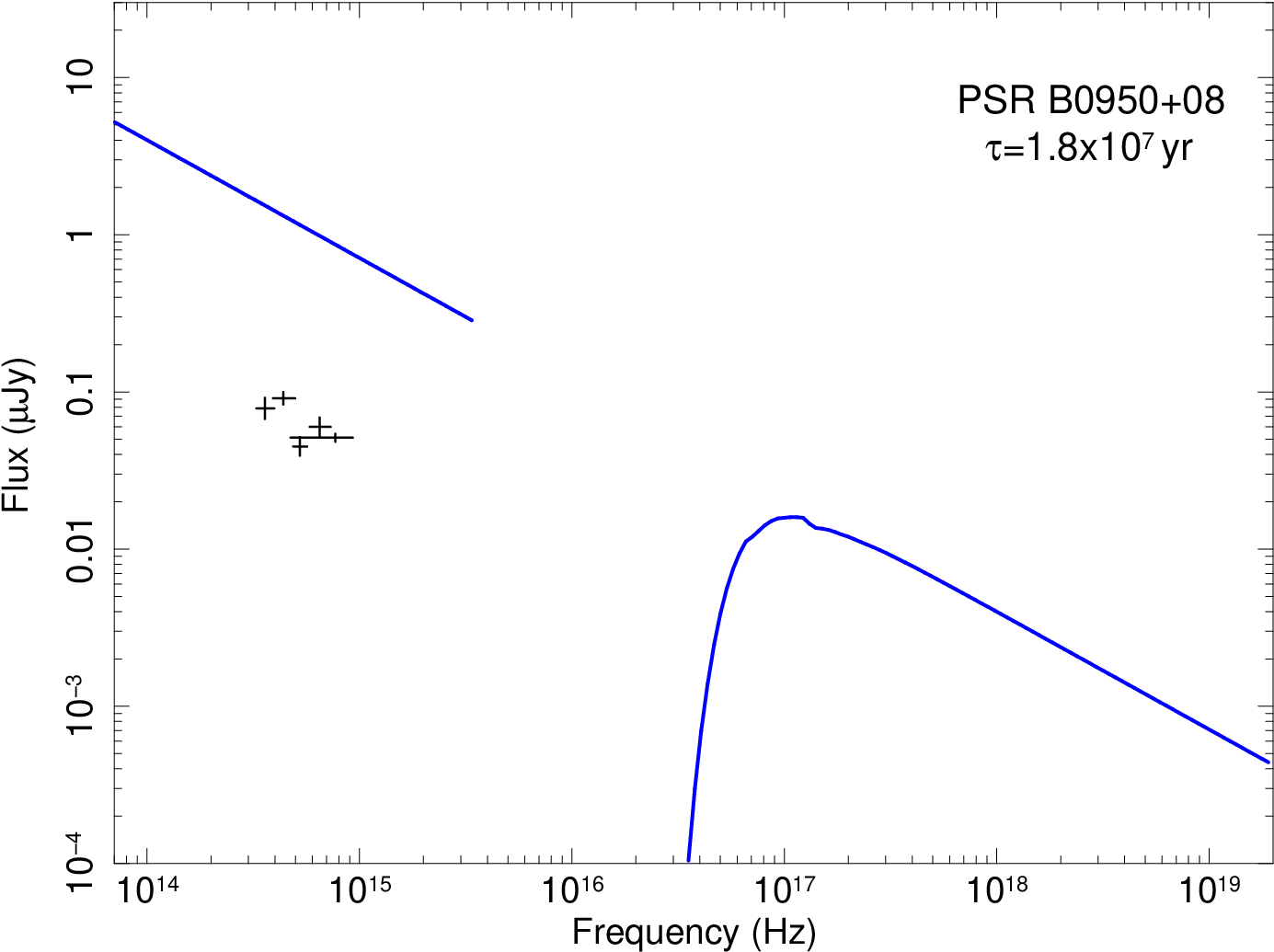}
  \caption{Same as Fig. \ref{multispec} but for all rotation-powered pulsars with an optical counterpart and flux measurements in at least two bands. Optical flux values are taken from the compilation  in Mignani et al. (2007a). X-ray spectral models are taken from the source publications: Crab (Willingale et al. 2001), PSR\, B1509$-$58 (Gaensler et al. 2002), Vela (Manzali et al. 2007), PSR\, B0656+14 (De Luca et al. 2005), Geminga (Caraveo et al. 2004),   PSR\, B1929+10 (Becker et al. 2006), PSR\, B0950+08 (Becker et al. 2004). Left panels: young pulsars   ($\tau < 10\,000$ years).    Right panels: middle-aged and old pulsars   ($\tau > 100\,000$ years).   In each column objects are sorted according to their spin-down age and labelled. Different X-ray spectral components are shown by the dotted red, green (black body), and blue (power-law) lines while the solid blue lines show the composite spectra.  Only best-fits are plotted for clarity.   }
\label{allspec}       
\end{figure*}

Spectral  turnovers between  these  two energy  bands  are a  common
feature of the magnetospheric emission of many other rotation-powered  pulsars, as indicated  by the
differences in their optical and X-ray spectral indexes (e.g., Mignani
et al. 2007a).  For  comparison,  we plotted  in
Fig. \ref{allspec} the optical,  near ultraviolet (NUV),  and near infrared (NIR) spectral fluxes for  all rotation-powered pulsars with  an optical counterpart, together with  the model  0.1-10 keV X-ray  spectra. In  the case of the Crab and Vela pulsars, the spectral turnover is in the  form  of a  single  break  in the  optical-to-X-ray power-law  spectrum. In  the case  of  \psr, instead,  a double break is required to join the optical and the X-ray SEDs, unless the actual interstellar extinction is larger by a factor of 2 with respect to current best estimates, which  would make the spectrum  flatten in the blue. This apparent double break in the optical-to-X-ray  power-law spectrum might be also present in  other, much closer, rotation-powered pulsars.   Of course, for  some pulsars
(especially for the  fainter ones) the comparison is  hampered by the
paucity of  spectral flux measurements at NIR/optical/NUV wavelengths,
which  makes it  difficult to  establish the  presence of  a power-law
component. Bearing this  caveat in mind, a possible double  break can be 
recognised  in the optical-to-X-ray power-law spectrum of  the middle-aged pulsars PSR\, B0656+14  ($\approx$100 kyrs) and, to a lesser extent, Geminga  ($\approx$ 340 kyrs),  for both of which a power-law tail is hinted in the NIR.  On the other hand, a single break is probably required for the old pulsars PSR\, B1929+10 ($\approx 3$ Myrs) and PSR\, B0950+08 ($\approx$ 18 Myrs), while only for the  young PSR\, B1509$-$58 the optical/NIR power-law seems to nicely  fit  the  extrapolation  of  the X-ray  one.  So,  three different trends  are recognised  in our sample  which spans  four age
decades. This means  that the breaks in the  power-law spectrum do not
correlate with the  neutron star age and, thus,  are not indicative of
any  evolution   in  the  emission  processes  in   the  neutron  star
magnetosphere.  This is  consistent with  the lack  of evidence  for a
power-law  spectrum  evolution,  both   in  the  optical  (Mignani  et
al.  2007a)  and  in  the  X-rays  (Becker  2009).  The shape of the power-law optical-to-X-ray spectrum does not correlate with  the dipole magnetic field either with, e.g. the Crab and PSR\, B0950+08 both featuring a single spectral  break but having magnetic fields different by  two orders  of magnitudes.  It is possible that  the number of observed spectral  breaks is related to  the geometry of the  emission regions and  to the particle
distribution  in the  neutron star  magnetosphere which,  however, are
difficult to reconstruct without  having information on both the X-ray
and  optical light  curve profiles.   At the same time, is  not possible  to
determine whether such double breaks are simply associated with a
change in the  power-law spectrum or they are associated with absorption
processes in the neutron star magnetosphere, a possibility proposed by
Serafimovich et al. (2004) for  \psr. Observations in the far UV (FUV)
would be  fundamental to clarify  this issue. Unfortunately,  only few
pulsars  have been  observed in  this  spectral region  by the  {\em  EUVE}
satellite (Korpela \&  Bowyer 1998) but only  Geminga (Bignami et  al. 1996) and
PSR\, B0656+14 (Edelstein et al.  2000) have been detected amongst the
pulsars shown in Fig.  \ref{allspec}, with their FUV fluxes being more or less compatible
with  the extrapolations  of  the black  body  components fitting  the
optical  spectra.

\subsection{The pulsar polarisation}

Our value of the phase-averaged polarisation ($PD = 16\%\pm 4\%$) is
consistent  with  the  upper  limit  of  $15  \%$  inferred  by
Middleditch  et  al. (1987)  from  phase-resolved  polarimetry of  the
pulsar. However,  the measured $PD$  is higher than  $PD \approx 5\%$
obtained from \vlt\  phase-averaged polarisation observations by Wagner \& Seifert (2000), for which, however, no  error estimate is given, so  that its significance
can not be  assessed.  Thus, ours is the  only significant measurement
($\sim 4  \sigma$ level)  of the  \psr\ phase-averaged
polarisation obtained so far.

Measurements  of  the  phase-averaged  polarisation degree  have  been
obtained  so  far  for  all   the  young  ($\la  10\,000$  years  old)
rotation-powered pulsars  with an  optical counterpart, albeit  with a
different  degree  of  confidence  (see  Mignani  et  al.   2007b  and
S\l{}owikowska et  al. 2009 for a  summary). For the Crab,  a value of
$PD=9.8\%  \pm  0.1\%$ was  inferred  from phase-resolved  polarimetry
observations  (S\l{}owikowska et  al. 2009),  while  image polarimetry
observations  with the \vlt\  yielded $PD=9.4\%\pm  4\%$ for  the Vela
pulsar (Mignani  et al.  2007b)  and $PD=10.4\%$ for  PSR\, B1509$-$58
(Wagner \& Seifert 2000), with the latter measurement being admittedly
very  uncertain   and  quoted  with  no   error.   The  phase-averaged
polarisation  of  \psr, $PD  =  16 \%\pm  4\%$,  is  thus  consistent with the measurements obtained for the other young pulsars,
as  one  might  expect  from  their similar parameters, like spin down luminosity and/or magnetic  field strength.
Although the  current data base  is still extremely limited,  with the
measurement obtained for PSR\, B1509$-$58 in wait for confirmation and
with  the phase-resolved polarimetry  observations of  the middle-aged
($\sim 100$ kyrs) PSR\, B0656+14  (Kern et al. 2003) covering only one
third of  the period, it nonetheless suggests  that the phase-averaged
optical polarisation  degree of rotation-powered  pulsars is typically
around $10\%$, with our value of $PD$ for PSR\, B0540$-$69 being possibly 
somewhat larger. 

On the theoretical side, the comparison of the observed phase-averaged polarisation degrees with the predictions of different pulsar magnetosphere models is complicated both by the degree of complexity of such models and by the limits in model simulations.  
However, for both the Crab and Vela pulsars, a detailed comparison with different pulsar magnetosphere  models, showed that the measured phase-averaged polarisation degree of the emerging optical radiation requires rather low degree of polarisation intrinsic to the source and/or strong depolarisation factors  (S\l{}owikowska
et al. 2009; Mignani et al. 2007b). Recent 3D outer-gap model calculations of optical and/or high-energy radiation with detailed incorporation of electron-positron gyration
have been carried out so far for the Crab pulsar only (Takata et al. 2007).
In general, the results seem to fit the measured optical polarization
level (see S\l{}owikowska et al. 2009) for some particular values of the viewing angle
(the angle between the line of sight and the spin axis). Other models addressing the problem of polarisation characteristics are relatively more simplified
in terms of the model assumptions in use  (Dyks et al. 2004; Petri \& Kirk 2005).
All models mentioned above - relevant for young energetic pulsars - rely on spatially extended sources of emission (either within the magnetosphere - Dyks et al 2004; Takata et al. 2007) or in the pulsar wind zone (Petri \& Kirk 2005). As a consequence, the depolarisation effects due to rotation, photon finite time of flight, and magnetic field structure are significant in all three models. This might explain the somewhat larger polarisation degree of PSR\, B0540$-$69. However, one should keep in mind that the shape of its optical light curve makes it  more difficult to infer the  geometry of the  emission region with respect to the Crab and Vela pulsars, which provides
crucial  input parameters  for  model simulations.  

\begin{figure}[h]
\centering
\includegraphics[height=8.5cm,angle=0,clip]{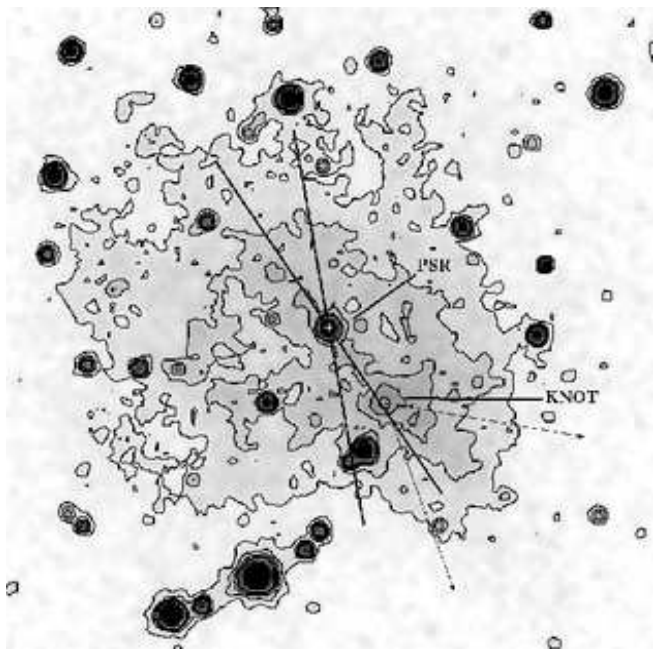}
  \caption{Same as in Fig. \ref{astro} but zoomed over a $8\arcsec \times 8\arcsec$ region.  \psr\  is labelled.  The image has been smoothed over $3\times3$ pixel cells for a better visualisation of the extended PWN emission. Isophotal contours (linear spacing) are drawn as solid lines. The thick lines show the measured $1 \sigma$ uncertainty ($\pm  12^{\circ}$) on the pulsar polarisation position angle, while the thin arrows show the assumed uncertainty ($\pm 30^{\circ}$) on the apparent proper motion  direction of the bright emission knot in the PWN (De Luca et al. 2007).  }
\label{pol}       
\end{figure}

Interestingly,  for both the  Crab and  Vela pulsars  the polarisation
position  angle  features a  remarkable  alignment  with  the axis  of
symmetry of the  X-ray structures (torus and jets)  observed by \chan,
with the pulsar spin axis,  and with the proper motion vector (S\l{}owikowska et al. 2009; Mignani et al.  2007b).   Unfortunately, for \psr\
the scenario is less clear since no proper motion has been measured so
far and \chan\ observations (Gotthelf \& Wang 2000; Petre et al. 2007)
could only partially  resolve the morphology of the  PWN, although its
clear asymmetry might hint at  the existence of a torus and, possibly,
of a jet.   A similar PWN morphology was also  observed in the optical
from \hst/\wfpc\ observations (Caraveo et al. 2000).  Most noticeably,
the \wfpc\ images shows the  existence of a bright emission knot south
west of the pulsar
which is  apparently moving at  a speed of  $\sim 0.04~c$ (De  Luca et
al. 2007) and  which can be interpreted as the head  of a possible jet
emitted by  the pulsar.   We note that  the orientation of  the pulsar
polarisation vector  ($PA =22^{\circ} \pm  12^{\circ}$) is consistent,
accounting  for   possible  projection  effects,  with   that  of  the
semi-major axis of the PWN (see Fig. \ref{pol}).  Moreover, the pulsar
phase-averaged polarisation  vector  is  also  possibly  aligned  with  the  apparent
direction  of motion  of  the knot  ($\approx  230^{\circ}$ east  from
north)\footnote{We note  that accounting for the proper  motion of the
LMC  (Costa  et al.  2009)  and for  the  galactic  rotation does  not
significantly  affect   this  apparent  alignment.}.   
We estimated the chance alignment probability  between the two vectors to be $\sim 0.04$, small but still not negligible.  If real, however, this  peculiar alignment would indicate a possible  physical connection between the pulsar and  the knot,  as already  proposed  in De  Luca et  al. (2007),  and
would support the pulsar/jet scenario. The measurement of both the knot and
of the PWN polarisation structure (Mignani  et al. in preparation) would further
reinforce this scenario.  Assuming for \psr\ the same  PWN geometry as
the Crab and Vela pulsars, the direction of the jet would suggest that
a putative torus would indeed  extend along the semi-minor axis of the
nebula, and  not along the semi-major one,  as previously hypothesised
from \chan\ and \hst\ observations  (Gotthelf \& Wang 2000; Caraveo et
al. 2000).

\section{Summary}

By using the \wfpc\ we performed a comprehensive study of the LMC pulsar \psr\ in the optical domain, including astrometry, multi-band photometry, and polarimetry, down to the limits achievable with the pre-refurbished \hst.

From the available multi-epoch images we obtained the tightest possible constraints on the \psr\ proper motion ($<$ 1 mas yr$^{-1}$) and transverse velocity ($<$ 250 km s$^{-1}$). Since the \wfpc\ has been now decommissioned and \psr\ has not been observed yet with \acs, it is unlikely that our limit will be improved by \hst\ before its retirement around 2015. Only the  {\em MICADO} adaptive optics IR camera at the {\em European Extremely Large Telescope}\footnote{http://www.eso.org/sci/facilities/eelt/} ({\em E-ELT}), with an expected sensitivity limit of $\sim 50 \mu$as yr$^{-1}$ (Trippe et al. 2010), will be eventually able to measure the pulsar proper motion. 
From the same \wfpc\  data set we measured the absolute position of \psr\  with an accuracy of $\sim 70$ mas, improving by a factor of 4 on the available X-ray coordinates.  A further improvement could be achieved by \hst\ using the refurbished \acs\ or the new {\em Wide Field Camera 3} ({\em WFC3}), which have a larger field-of-view and a smaller pixel scale with respect to the \wfpc\ mosaic, and the future {\em Gaia} astrometric catalogue which, however, will not be released in its final version until  2019. 
From observations in six different filters we characterised the \psr\ optical SED,  as accurately as it could be done  through broad-band imaging photometry with the \hst, confirming that the pulsar spectrum is a modelled by a power-law ($\alpha_{O}=0.70 \pm 0.07$) without evidence of a previously claimed spectral turnover in the U band. While multi-band photometry with  \acs\ and {\em WFC3} can certainly improve on our result,  especially in the red and blue part of the spectrum, high-spatial resolution spectroscopy with either of the two instruments or with the refurbished {\em Space Telescope Imaging Spectrograph} ({\em STIS}), is the only way to obtain a more detailed characterisation of the pulsar optical SED. 
Finally, from  image polarimetry we  derived a new measurement  of the
phase-averaged polarisation  degree of \psr\  ($PD = 16\%  \pm 4\%$;
$PA=22^{\circ}   \pm  12^{\circ}$),   with  the   polarisation  vector
apparently aligned with the PWN semi-major axis and with the apparent
proper  motion direction  of  its  bright emission  knot  (De Luca  et
al. 2007).  Multi-band polarimetry observations  with the refurbished,
higher  throughput, \acs\  will allow  to obtain  both a  more precise
measurement of  the pulsar  phase-averaged polarisation degree and of its positional angle,  and to
investigate a possible dependence  on the wavelength.  
Phase-resolved polarisation measurements in the optical, as well as in the  X and $\gamma$-rays, are crucial for  testing and constraining current models of pulsar
magnetosphere activity. Unfortunately, sensitive fast high-energy polarimeters, like eg. {\em PoGOLite} (Kamae et al. 2008),  are still in the phase of developments, and they will not be available soon. For the time being the phase-resolved polarimetric studies in optical remain the only way to carry out such investigations.
Therefore, phase-resolved optical polarimetry which, at present, can only be performed with guest instrument at  ground-based telescopes, like {\em  Optima} (Kanbach et
al. 2009) and  {\em GASP} (Collins et al. 2009),  will be then unique and crucial
to study the polarisation properties of noncoherent magnetospheric
radiation from PSR\, B0540$-$69 and other optical pulsars.  In particular, given the  faintness of \psr\ in the  radio band, polarisation
measurements in the optical domain are  the only way to carry out such
investigations.

About 25 years after its discovery, \psr\ is still the only extra galactic optical pulsar identified so far.  Thus, it is a unique target for optical observations with the present ground and space-based facilities.  More extra galactic neutron stars, however, are expected to be discovered in the optical, both in the Magellanic Clouds and beyond, after the advent of the {\em E-ELT} which holds the potential of detecting Crab-like optical pulsars up to the distance of M31 (Shearer 2008).

\begin{acknowledgements}
RPM  acknowledges  STFC  for  support  through  a  Rolling  Grant and thanks G. Soutchkova (STScI) for support in the observation scheduling and S. Bagnulo and J. Dyks  for useful discussions. The authors thank S. Zharikov for arranging in tabular format the data plotted in Fig. \ref{allspec}. A.  S\l{}owikowska\ is partially supported by the European Union Marie Curie grant MTKD-CT-2006-039965 and Polish Grant N N203 2738 33.
\end{acknowledgements}

\end{document}